\documentclass{PoS}
\usepackage{psfig,epsfig,colordvi,amssymb,amsmath}

\usepackage{slashed}

\def\lsim{\mathrel{\rlap{\lower4pt\hbox{\hskip1pt$\sim$}}
    \raise1pt\hbox{$<$}}}                
\def\gsim{\mathrel{\rlap{\lower4pt\hbox{\hskip1pt$\sim$}}
    \raise1pt\hbox{$>$}}}                

\newcommand{\be}{\begin{equation}}
\newcommand{\ee}{\end{equation}}
\newcommand{\bea}{\begin{eqnarray}} 
\newcommand{\eea}{\end{eqnarray}}

\newcommand{\Dlr}{\mbox{\parbox[b]{0cm}{$D$}\raisebox{1.7ex}
                       {${\,\scriptstyle{\leftrightarrow}}$}}}

\newcommand{\Op}{\mathcal{O}} 

\title{Nucleon form factors and moments of parton
distributions in twisted mass lattice QCD}

\ShortTitle{Nucleon generalized form factors using TMF}

\author{C. Alexandrou~$^{(a,b)}$, \speaker{M. Constantinou}~$^{(a)}$, C. Kallidonis~$^{(a)}$, T. Korzec~$^{(a,c)}$\\
$^{(a)}$ Department of Physics, University of Cyprus, P.O. Box 20537, 1678 Nicosia, Cyprus\\
$^{(b)}$ Computation-based Science and Technology Research
    Center, Cyprus Institute, 20 Kavafi Str., Nicosia 2121, Cyprus    \\
$^{(c)}$ Institut f\"ur Physik
   Humboldt Universit\"at zu Berlin, Newtonstrasse 15, 12489 Berlin, Germany\\
        E-mail: \email{alexand@ucy.ac.cy},
        \email{constantinou.martha@ucy.ac.cy}, \email{kallidonis.christos@ucy.ac.cy}, \email{korzec@physik.hu-berlin.de}}
\author{M. Brinet, J.~Carbonell,  P.~A.~Harraud, M.~Papinutto\\
 Laboratoire de Physique Subatomique et Cosmologie,
               UJF/CNRS/IN2P3, 53 avenue des Martyrs, 38026 Grenoble, France\\
E-mail:\email{mariane@lpsc.in2p3.fr}, \email{Jaume.Carbonell@lpsc.in2p3.fr}, \email{harraud@lpsc.in2p3.fr}, \email{Mauro.Papinutto@lpsc.in2p3.fr}}
\author{P. Guichon\\
CEA-Saclay, IRFU/SPhN, 91191 Gif-sur-Yvette, France\\
        E-mail: \email{pierre.guichon@cea.fr}}
\author{K. Jansen\\
NIC, DESY, Platanenallee 6, D-15738 Zeuthen, Germany\\
        E-mail: \email{Karl.Jansen@desy.de}}

\abstract{
We present results on the electroweak form factors
and on the lower moments of parton
distributions of the nucleon, within lattice QCD using two dynamical flavors
of degenerate twisted mass fermions. Results are obtained on
lattices with three different values of the lattice spacings, namely
a=0.089~fm, a=0.070~fm and a=0.056~fm, allowing the
investigation of cut-off effects. The volume dependence is examined
by comparing results on two lattices of spatial length L=2.1~fm and
L=2.8~fm. The simulations span pion masses in the range of 260-470~MeV.
Our results are renormalized non-perturbatively and the values
are given in the MS-scheme at a scale $\mu$=2 GeV.
}

\FullConference{XXIst International Europhysics Conference on High Energy Physics\\
                21-27 July 2011\\
                Grenoble, Rhones Alpes France}

\begin{document}

\vspace{-0.25cm}
\section{Introduction}
\vspace{-0.15cm}
Lattice QCD calculations of observables related to the
structure of baryons are now being carried out using simulations of the
theory with parameters close to the
physical regime both in terms of pion mass and with respect to the
continuum limit~\cite{Hagler:2007xi, Brommel:2007sb, Alexandrou:2010cm, Yamazaki:2009zq, Alexandrou:2010hf}. 
In particular, a number of major collaborations are investigating the form
factors and the first moments of parton distributions of the nucleon
that encode important information on its structure.
In this work we consider 
 the nucleon matrix elements of the 
operators 
$
   \Op_V^{\mu_1\ldots\mu_{n}}    =
  \bar \psi  \gamma^{\{\mu_1}i\Dlr^{\mu_2}\ldots i\Dlr^{\mu_{n}\}}
  \psi$ and
$\Op_{A}^{\mu_1\ldots\mu_{n}}     =
\bar \psi  \gamma^{\{\mu_1}i\Dlr^{\mu_2}\ldots
i\Dlr^{\mu_{n}\}} \gamma_5\psi   \, 
$
limiting ourselves to $n=1$ and $n=2$.
The case $n=1$
reduces to 
the nucleon form factors. The nucleon matrix element of the electromagnetic current 
 is written in the form 
\vspace{-0.25cm}
\be 
 \langle \; N (p',s') \; | j^\mu | \; N (p,s) \rangle =   
  \biggl(\frac{ m_N^2}{E_{N}({\bf p}^\prime)\;E_N({\bf p})}\biggr)^{1/2}   
  \bar{u}_N (p',s') \left[ \gamma^\mu F_1(q^2)   
+  \frac{i\sigma^{\mu\nu}q_\nu}{2m_N} F_2(q^2)\right] u_N(p,s) \; ,  
\label{NjN}  
\ee  
\vspace{-0.15cm}
where $q^2=(p^\prime-p)^2$, 
$ m_N$ is the nucleon mass and $E_N({\bf p})$ its energy. 
$F_1(0)=1$ for the proton and zero for the neutron since we have a conserved current. 
$F_2(0)$ measures the anomalous magnetic moment. They are connected to the 
electric, $G_E$, and magnetic, $G_M$, Sachs form factors by the relations: 
$ 
G_E(q^2)= F_1(q^2) + \frac{q^2}{(2m_N)^2} F_2(q^2)$ and
$G_M(q^2)= F_1(q^2) + F_2(q^2). 
$ 
Similarly the matrix element for the axial-vector current is given by 
\vspace{-0.25cm}
\be
\langle N(p',s')|A_\mu^3|N(p,s)\rangle= i \Bigg(\frac{
            m_N^2}{E_N({\bf p}')E_N({\bf p})}\Bigg)^{1/2} 
            \bar{u}_N(p',s') \Bigg[
            G_A(q^2)\gamma_\mu\gamma_5 
            +\frac{q_\mu \gamma_5}{2m_N}G_p(q^2) \Bigg]\frac{1}{2}u_N(p,s).
\label{axial ff}
\ee
\vspace{-0.15cm}
 We also study matrix
elements of operators with a single derivative ($n=2$). 
The matrix elements of these operators are parameterized in terms
of the generalized form factors
$A_{20}(Q^2),\ B_{20}(Q^2),\ C_{20}(Q^2)$ and $\tilde
A_{20}(Q^2),\ \tilde B_{20}(Q^2)$ ($Q^2=-q^2$). The results 
are non-perturbatively renormalized~\cite{ACKPS}.
In this work we use simulations with $N_F=2$  twisted mass
fermions (TMF). They provide an attractive formulation of
lattice QCD that allows for automatic ${\cal O}(a)$ improvement,
infrared regularization of small eigenvalues and fast dynamical
simulations~\cite{Shindler:2007vp}. 
For the calculation of the nucleon observables discussed in this work the
automatic ${\cal O}(a)$ improvement is particularly relevant since it
is achieved by tuning only one parameter in the action, requiring no further improvements on the
operator level.
Details on the lattice evaluation can be found in 
 Refs.~\cite{Alexandrou:2010hf}.

\vspace{-0.25cm}
\section{Nucleon form factors}
\begin{figure}
\begin{center}
   \includegraphics[scale=0.295]{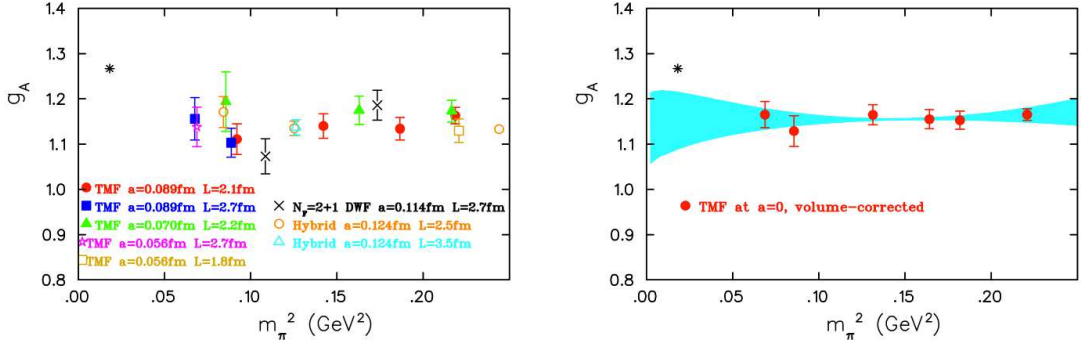}
\end{center}
\vspace{-0.5cm}
   \caption{\label{fig:axial_charge} Left: Results on $g_A$ as a
     function of the pion mass squared using TMF, DWF and domain wall valence on a staggered sea (hybrid). Right: $g_A$ obtained by taking the
     continuum limit of the volume corrected TMF data. The shaded area
     is the best chiral fit to the data shown on the graph. }
\end{figure}

\vspace{-0.15cm}
Our lattice results on the nucleon axial charge, $g_A$
are shown in Fig.~\ref{fig:axial_charge}.
We observe that results at our three different lattice spacings are
within error bars and that results at the two different volumes are
also consistent. In the same figure we compare 
 to the results obtained using $N_F=2+1$ 
domain wall fermions (DWF) by the RBC-UKQCD collaborations
\cite{Yamazaki:2009zq} and using a hybrid action with 2+1 flavors of
staggered sea and domain wall valence fermions by LHPC~\cite{Bratt:2010jn}.
There is agreement among lattice results
using different lattice actions even before taking the continuum and
infinite volume limit indicating that lattice artifacts are small.
\begin{figure}
\begin{center}
\begin{minipage}{0.42\linewidth}
      {\includegraphics[width=\linewidth]{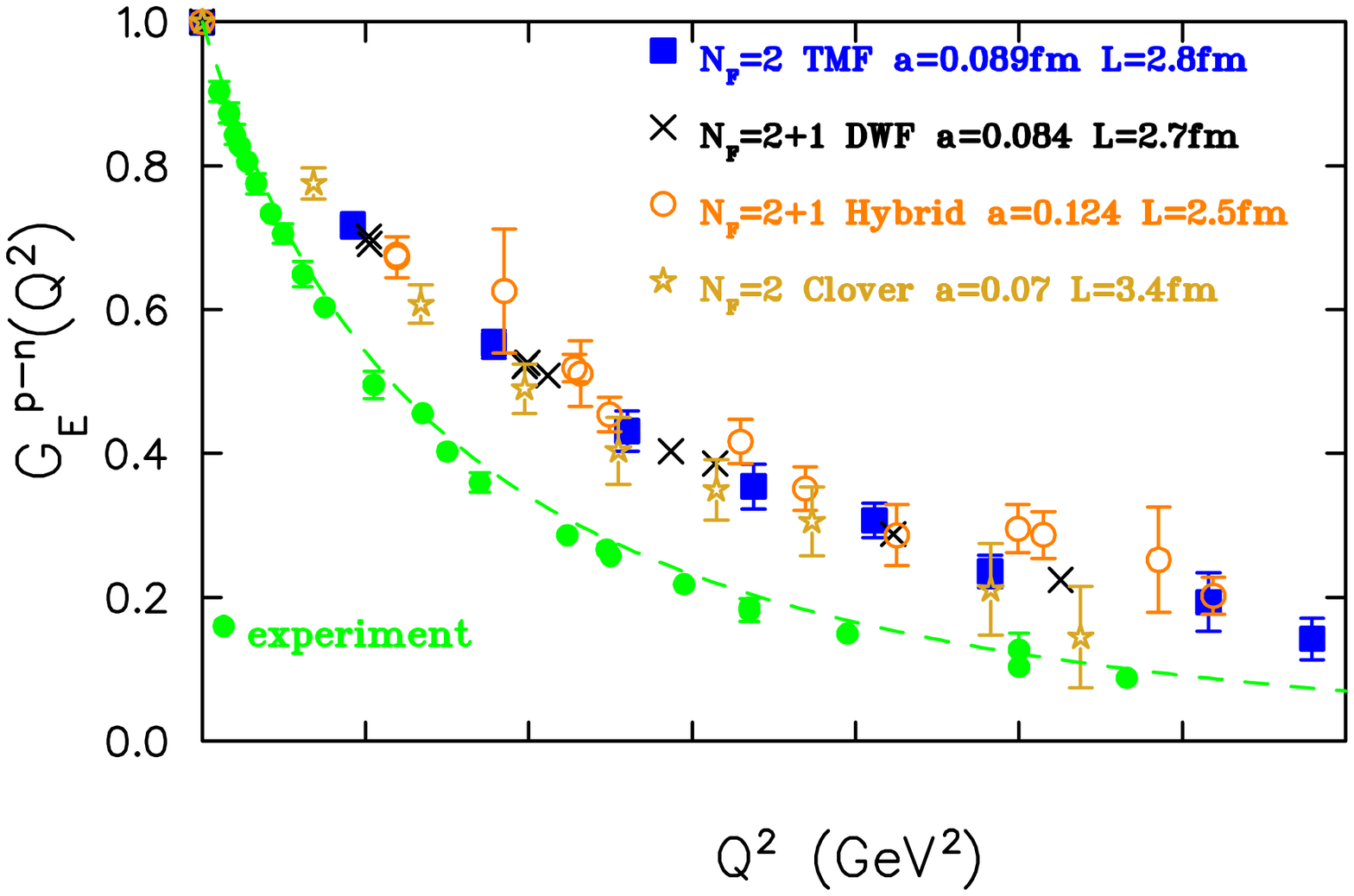}}\vspace{-0.05cm}
      \hspace{0.1cm}{\includegraphics[width=\linewidth]{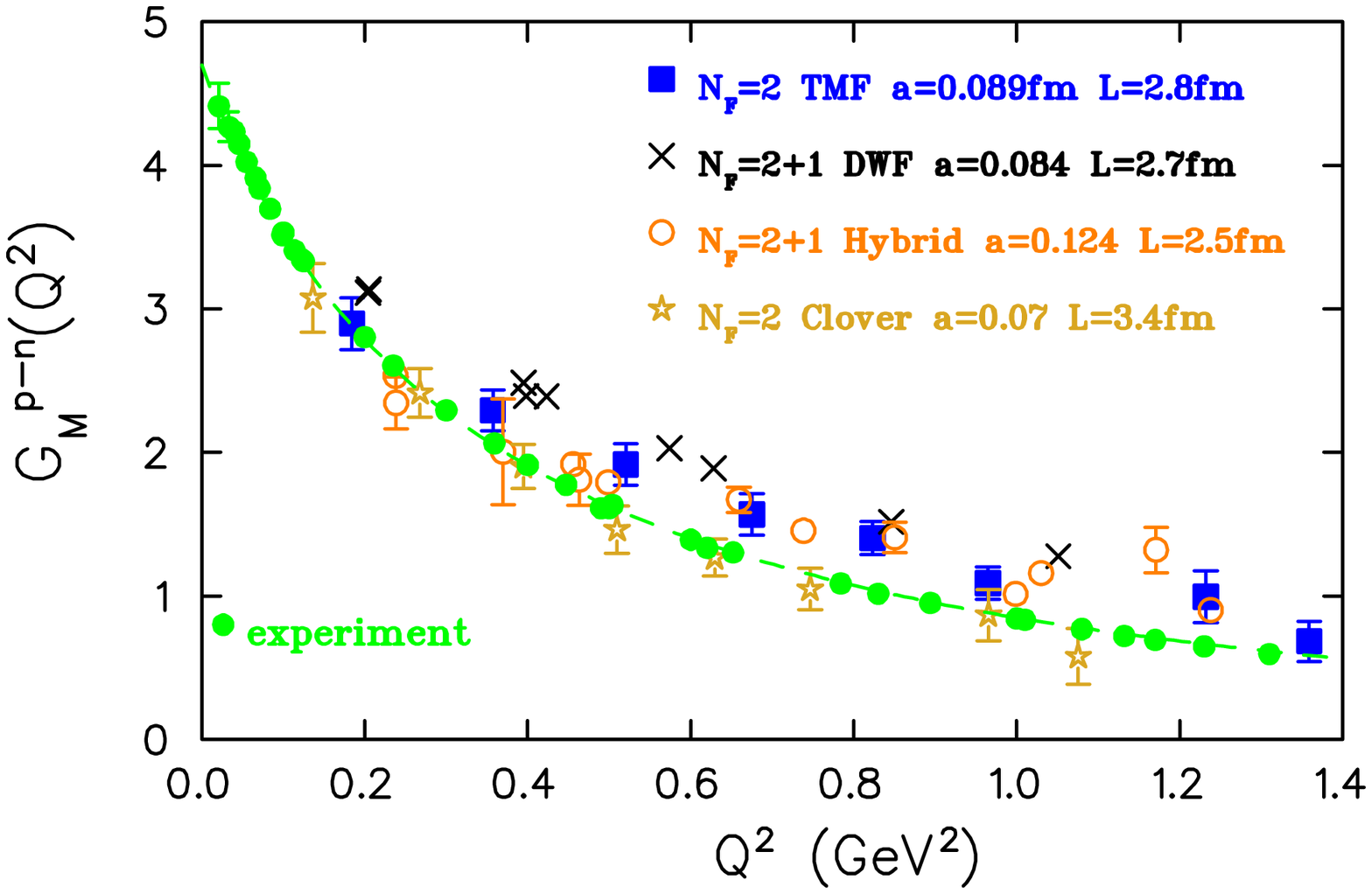}}
   \end{minipage}
\begin{minipage}{0.42\linewidth}
   {\includegraphics[width=\linewidth]{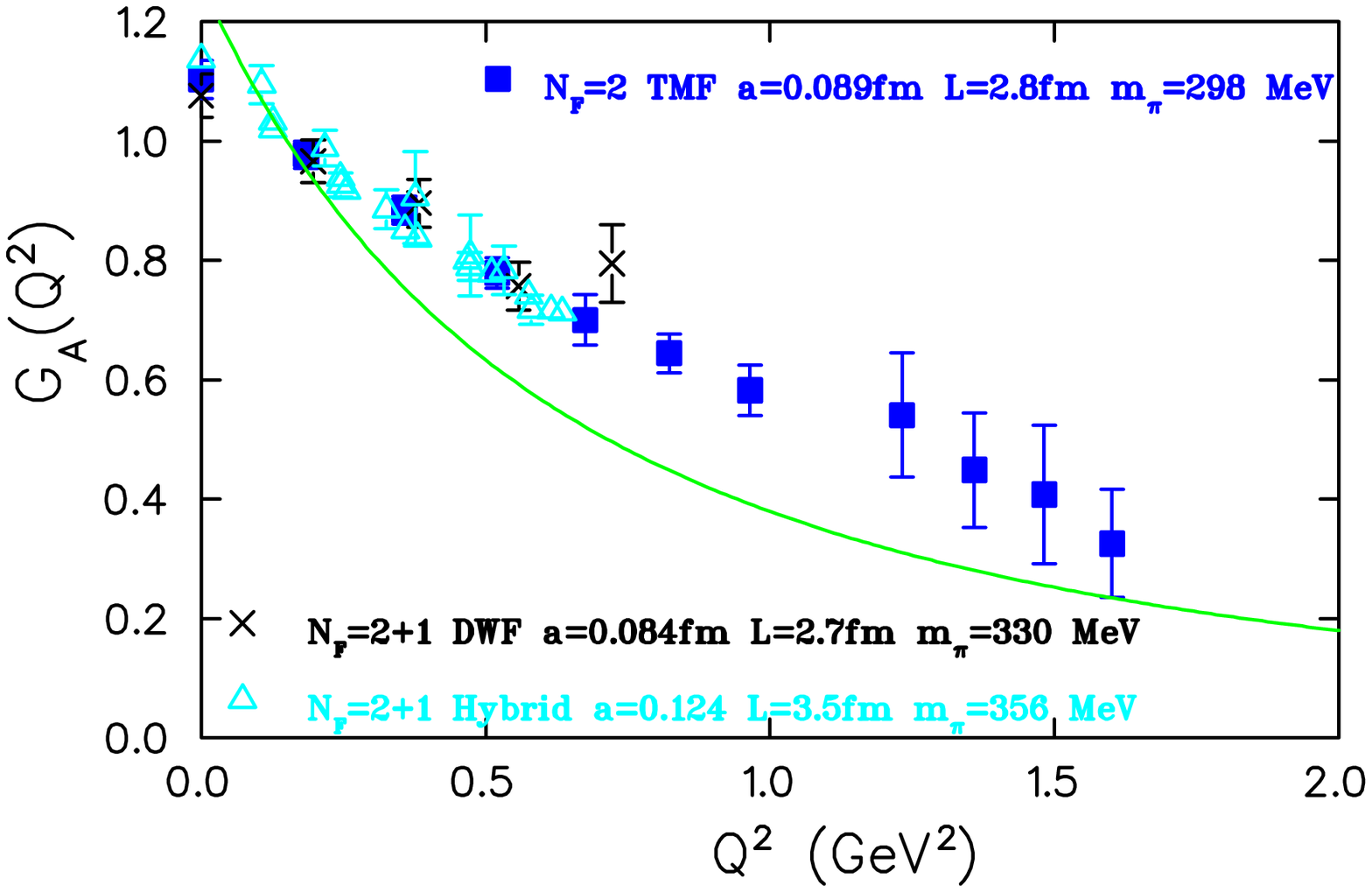}}\vspace{-0.05cm}
     {\includegraphics[width=\linewidth]{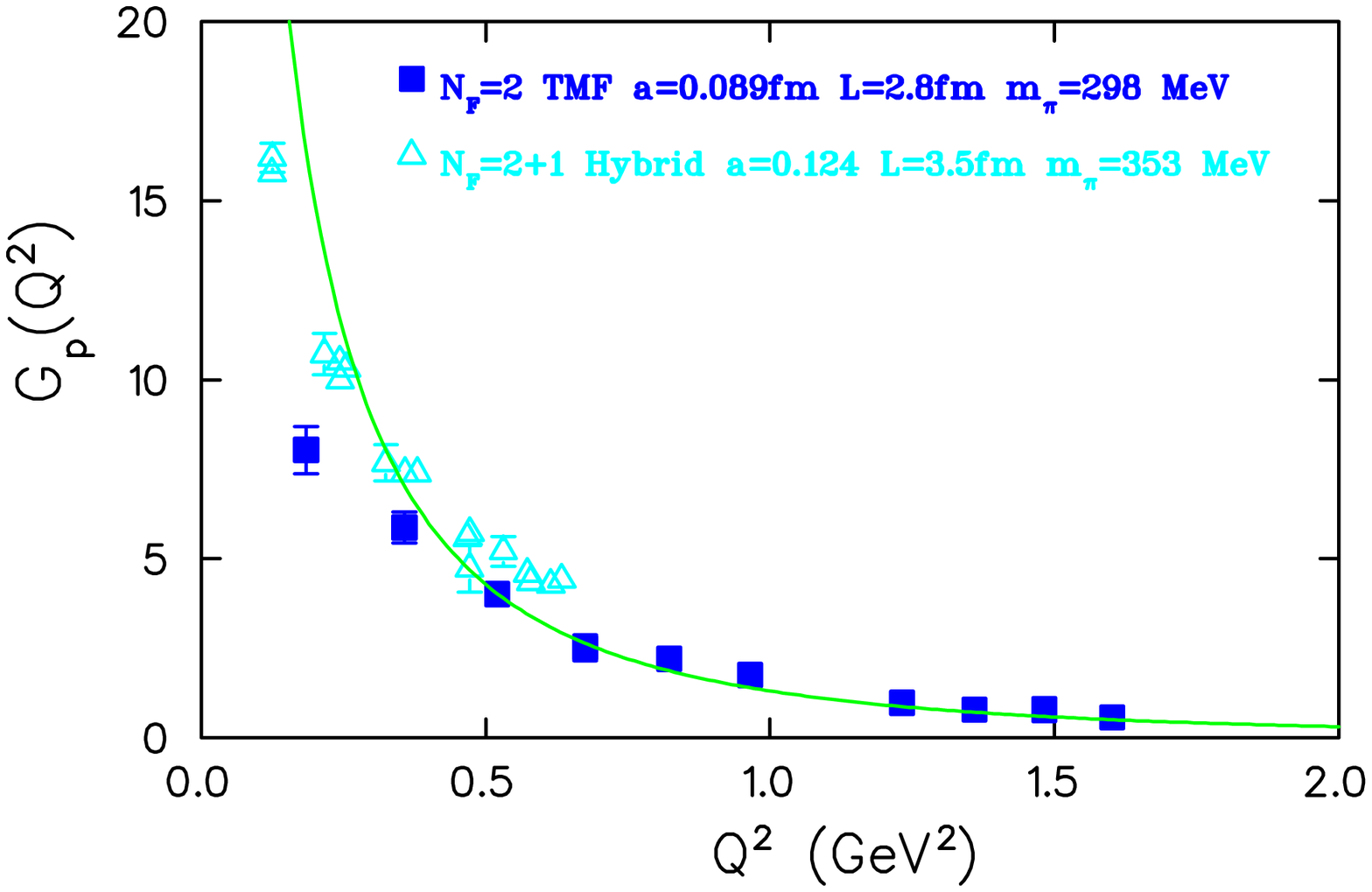}}
   \end{minipage}
\caption{\label{GE_Gp} $G_E$ and $G_M$ at $m_\pi\sim$300~MeV (left) and $G_A$ and $G_p$ (right)  for various lattice actions. 
  The filled (green) circles show experimental data on $G_E$ 
and $G_M$. The solid line on the
 right panel is a dipole fit to experimental data for $G_A(Q^2)$
combined with pion pole dominance to get the solid curve shown for $G_p(Q^2)$.}
\end{center}
\vspace*{-1cm}
\end{figure}
The pion mass dependence for the nucleon axial
charge has been studied within HB$\chi$PT in the SSE
formulation~\cite{Hemmert:2003cb}.
Fitting the volume corrected and continuum extrapolated TMF results
shown in Fig.~\ref{fig:axial_charge} we find $g_A{=}1.12(8)$, while fitting the raw lattice results
 we obtained $g_A{=}1.08(8)$ confirming that cut-off and volume
corrections have indeed a small effect.

In Fig.~\ref{GE_Gp} we show the isovector electric and magnetic
 form factors at $m_\pi\sim 300 $~MeV as a
function of the momentum transfer squared. We compare our results with
 results using other ${\cal O}(a)$ improved actions at the same pion mass~\cite{Bratt:2010jn,Capitani:2010sg,Syritsyn:2009mx}. 
The results are in agreement in the case of $G_E$, while in the
case of $G_M$ clover results from Ref.~\cite{Capitani:2010sg} are lower.
In the same figure we also show TMF results on the two axial-vector form
factors and we compare with DWF at $m_\pi=330$~MeV~\cite{Yamazaki:2009zq} and hybrid results 
by LHPC at $m_\pi=356$~MeV~\cite{Bratt:2010jn} on a lattice
with $L=3.5$~fm. The results are in  agreement in the case of $G_A(Q^2)$,
while in the 
case of $G_p(Q^2)$ there are  discrepancies at low $Q^2$ values, which
 may indicate that volume effects are not negligible 
on form factors such as 
$G_p(Q^2)$ which are strongly affected by the pion-pole.

\vspace{-0.25cm}
\section{Nucleon moments}
\vspace{-0.15cm}
\begin{figure}
{\includegraphics[scale=0.43]{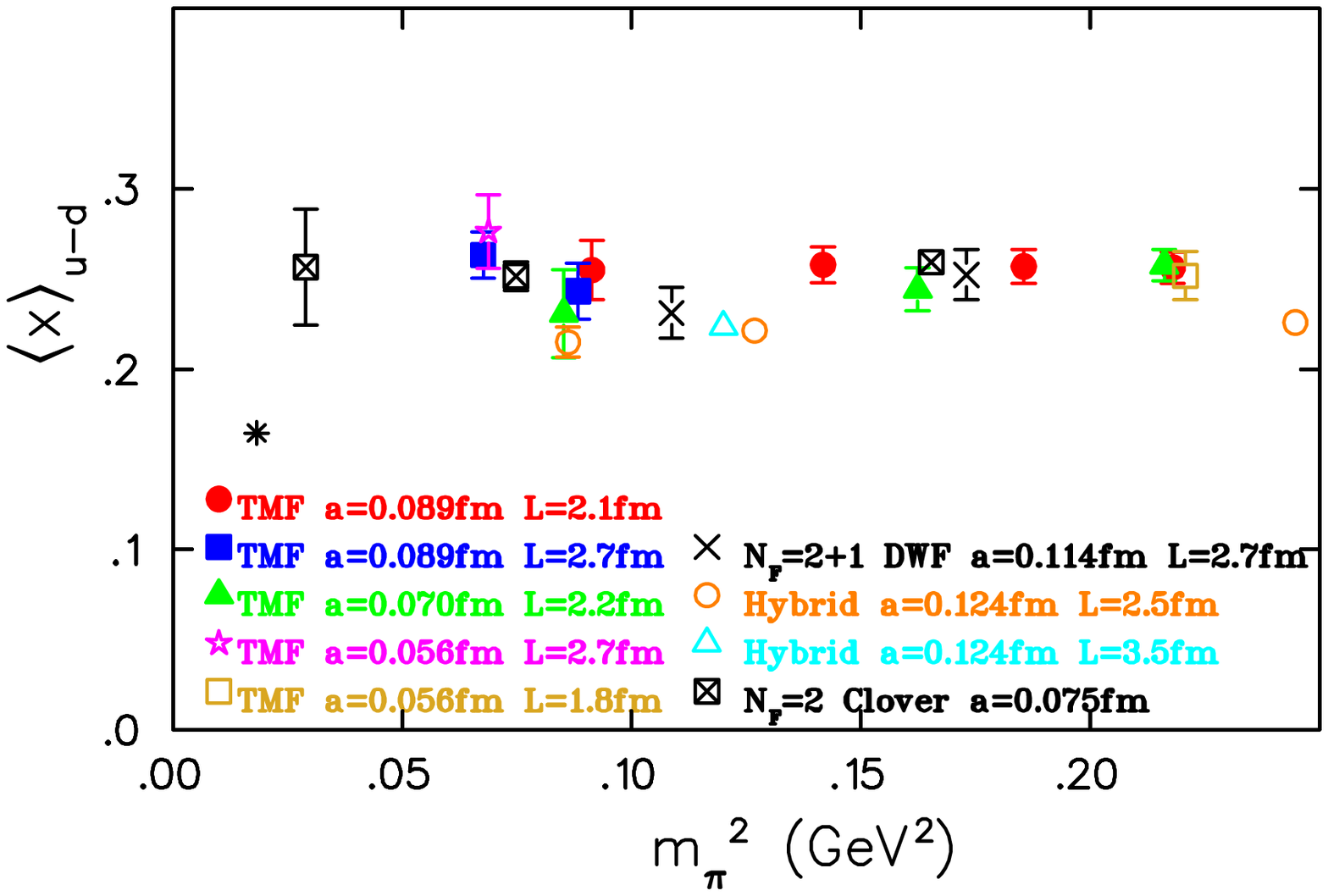}}$\quad$
{\includegraphics[scale=0.43]{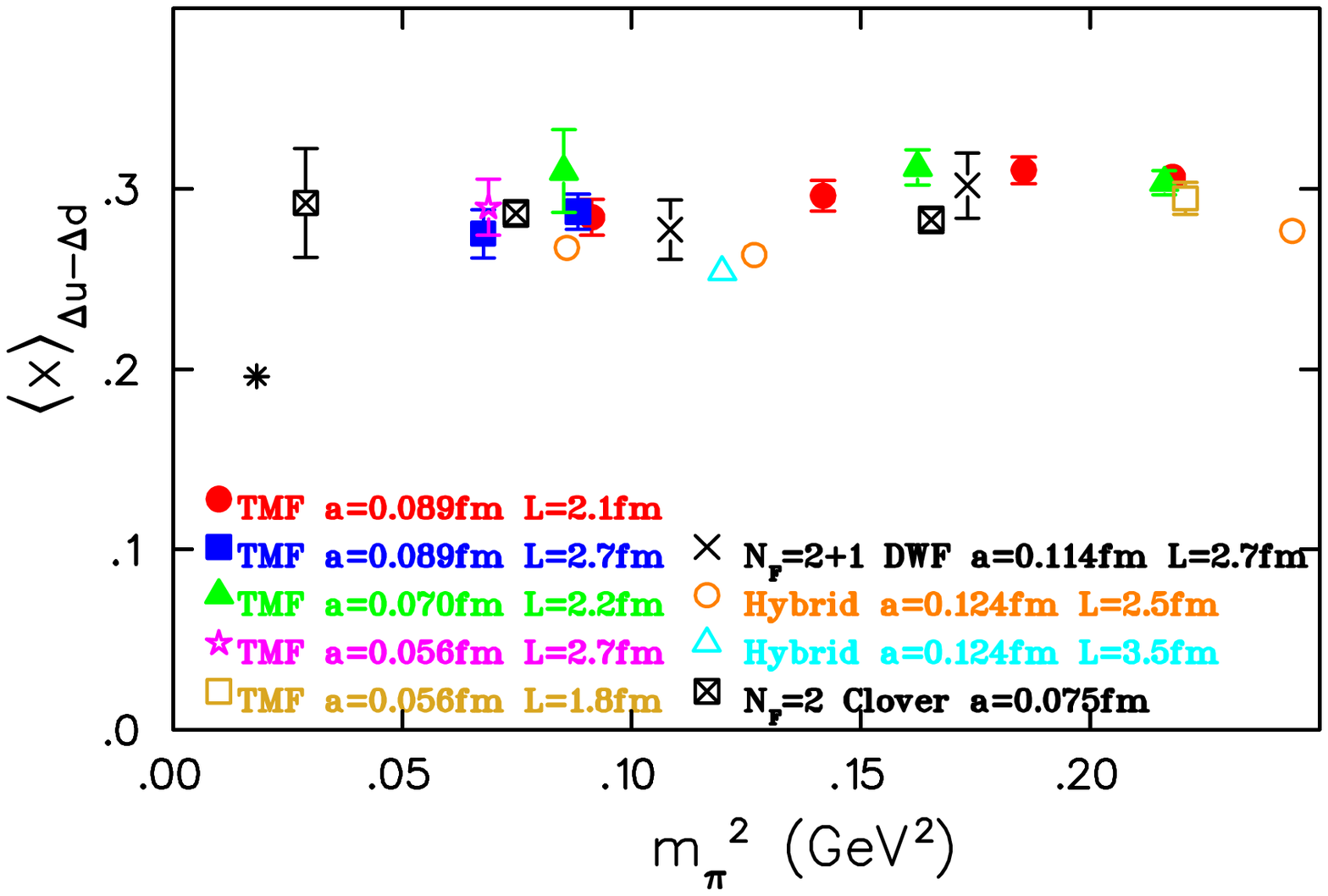}}
\caption{Our lattice data on $\langle x\rangle _{u-d}$ ($A_{20}$)
 and $\langle x\rangle_{\Delta u-\Delta d}$ ($\tilde{A}_{20}$) as a
 function of the pion mass squared. Results using other lattice
 actions are also plotted. The physical point is shown by the asterisk.}
\label{fig:compare}
\vspace*{-0.3cm}
\end{figure}

In Fig.~\ref{fig:compare} we show our results on the spin-independent and helicity
moments, and results using other lattice actions~\cite{Bratt:2010jn,Pleiter:2011gw,Aoki:2010xg}.
Although to compare lattice data using different discretization
schemes one would have to first extrapolate to the continuum limit, we
find a good agreement among lattice results, due to small cut-off
effects for lattice spacings of about 0.1~fm. Lattice values for
$\langle x \rangle_{u-d} =A_{20}(Q^2=0)$ although compatible,
are higher from the phenomenological value $\langle x_{u-d} \rangle
\sim 0.16$. A similar conclusion holds for the helicity moment.

\vspace{0.25cm}
{\bf{Acknowledgments:}}
This work was partly supported by funding received from the
Cyprus Research Promotion Foundation under contracts EPYAN/0506/08,
and TECHNOLOGY/$\Theta$E$\Pi$I$\Sigma$/0308(BE)/17.
M. P. acknowledges financial support by a Marie Curie European Reintegration
Grant of the 7th European Community Framework Programme under contract
number PERG05-GA-2009-249309. 
\vspace*{-0.3cm}
%
%

\end{document}